# A Gmail-Based Phishing Detection Prototype for Nigerian Fintech Emails Using Sender Checks and BiLSTM Classification

Francis Gideon Oghie[1], Uche Emmanuel Unoke[1]

[1]Department of Cyber Security Science, School of Information and Communication Technology, Federal University of Technology, Minna, Nigeria

**Abstract**

Phishing emails that impersonate Nigerian fintech providers can combine deceptive sender addresses, lookalike links, and locally familiar language. This study presents a Gmail browser extension that integrates sender-domain and URL checks with a bidirectional long short-term memory (BiLSTM) classifier. The extension compares visible sender addresses and links with profiles for eight fintech platforms, obtains a phishing probability from a locally hosted Flask service, and displays a legitimate, warning, or phishing verdict when an email is opened.

The BiLSTM classifier was evaluated on 8,943 test messages from a cleaned dataset of 59,622 phishing and legitimate emails. The test confusion matrix recorded 4,308 true negatives, no false positives, one false negative, and 4,634 true positives. These counts correspond to 99.99% accuracy, 100.00% precision, 99.98% recall, and 99.99% F1 score. Tokenized sequence analysis identified 5.79% overlap between the training and test sets, which may inflate performance estimates for independent messages. A Gmail demonstration showed the integrated extension producing user-visible verdicts, although the complete system was not evaluated on a labeled test set. The findings establish the feasibility of the implemented prototype while leaving its end-to-end detection performance and generalization to unseen attacks open for further evaluation.

## 1. Introduction

Email remains a channel through which financial service providers communicate transaction updates, account notices, and security alerts. Phishing attacks exploit these familiar messages by impersonating providers and directing recipients to fraudulent websites or requests for sensitive information. In Nigeria, an email that appears to come from a fintech platform may combine a recognizable brand name with a lookalike sender domain, a deceptive link, and language familiar to its intended recipient. Each element can affect whether the message appears credible when it is opened.

Phishing email detection has been studied using message features, machine learning, and natural-language processing. Fette et al. demonstrated that features associated with deceptive email can support automated classification [1]. Subsequent research applied long short-term memory (LSTM) networks to email data [2] and combined textual and other email features in hybrid detection methods [3]. These approaches establish useful detection techniques, but their reported results depend on the messages and evaluation procedures used in each study. Their performance alone does not establish how a detector will behave when emails impersonate particular Nigerian fintech providers.

A detector for this setting must handle several distinct signals. Suspicious wording may indicate an attempt to create urgency, while a sender domain or embedded URL may resemble an official fintech address without matching it. Content classification cannot determine whether a sender belongs to a known provider, and domain matching alone cannot assess the message's language. Combining these signals also introduces a design question: how should the system respond when its checks disagree?

This study presents an implemented phishing detection prototype for Gmail that addresses these questions through three components. A browser extension extracts the visible sender address, message text, and links when an email is opened. Sender and URL checks compare addresses and destinations with locally stored profiles for eight Nigerian fintech platforms and identify possible lookalike domains. A bidirectional LSTM (BiLSTM) classifier assigns a phishing probability to the message text through a local Flask inference service. Decision rules combine the outputs and display a legitimate, warning, or phishing verdict within Gmail. The sender check is based on the address visible in the interface; it does not perform cryptographic email authentication.

The study's contribution is the design and implementation of this integrated workflow for Nigerian fintech email, together with an evaluation of its text classifier. The BiLSTM was tested on 8,943 messages, and the complete extension was demonstrated in Gmail. The classifier results and the extension demonstration answer different questions: the former measures text classification under the recorded test conditions, while the latter shows that the components operated together in the browser. Interpretation of the classifier result must also account for overlap identified between the training and test data. These boundaries define what can be concluded about the prototype and what remains to be established through an independent, end-to-end evaluation.

## 2. Related Work

Phishing email detection has developed from manually defined indicators toward models that learn patterns from labeled messages. Fette et al. [1] used features associated with deceptive email to distinguish phishing from legitimate messages. Their work established the value of examining characteristics of the message beyond isolated suspicious words. Its evaluation also illustrates why a detection result must be read in the context of the dataset on which it was obtained.

Later studies investigated whether models could learn patterns directly from email content. Li et al. [2] applied long short-term memory networks to phishing detection in large email datasets, using the model's ability to process sequences of text. Alhogail and Alsabih [4] examined natural-language processing and a graph convolutional approach to classifying phishing emails from message content. These studies support text classification as one component of a detector, while leaving sender identity and link destinations to be addressed through other signals.

Research has also combined different types of evidence within a single detection method. Bountakas and Xenakis [3] proposed HELPHED, which applies ensemble learning to textual and content-based email features. Adeyemo et al. [5] combined email text and URL features in a phishing detection model. Both studies demonstrate the relevance of multiple signals, although their architectures and evaluation datasets differ from the Gmail extension examined here. Their reported performance therefore cannot serve as a direct baseline for this study.

Sender information introduces a separate issue: a familiar address displayed to the recipient does not, by itself, establish that the message originated from the organization it names. DMARC evaluates alignment between the domain in an email's visible author address and an identifier authenticated through SPF or DKIM [6]. The present system uses a local directory to compare the sender address displayed in Gmail with known fintech profiles. That comparison can contribute to phishing detection, but it is distinct from the authentication checks defined by DMARC.

The present study builds on established text-classification and feature-combination approaches by applying them in an implemented Gmail workflow focused on Nigerian fintech impersonation. Its distinguishing element is the integration of a BiLSTM classifier with platform-specific sender profiles, URL checks, and user-visible verdicts. The evaluation reported here measures the BiLSTM classifier and demonstrates operation of the integrated extension. It does not measure the separate contribution of each component or establish a performance advantage over the systems reviewed above.

## 3. Methodology and System Design

### Data preparation

The email corpus combined publicly available samples, legitimate Nigerian fintech emails, and locally constructed phishing examples. Approximately 1,000 phishing examples were created before cleaning to represent platform impersonation, lookalike domains, and Nigerian English and Pidgin expressions. Source labels were not retained throughout preprocessing, so the number of constructed examples in the final corpus cannot be determined.

The initial collection contained 66,032 records. Email addresses, phone numbers, and account numbers were redacted; HTML was stripped; duplicates were removed using content hashes; and content outside the English and Nigerian Pidgin scope was excluded. The resulting dataset contained 59,622 emails: 30,929 labelled phishing and 28,693 labelled legitimate. Stratified sampling allocated 41,736 emails to training, 8,943 to validation, and 8,943 to testing. Nigerian Pidgin terms were retained during text preparation.

**Text classification model**

Email text was converted to token sequences using a vocabulary of 10,000 terms. Each sequence was padded or truncated to 500 tokens. An embedding layer initialized with 100-dimensional GloVe vectors fed two bidirectional long short-term memory (BiLSTM) layers with 128 and 64 units, respectively. A sigmoid output unit produced a phishing probability between zero and one.

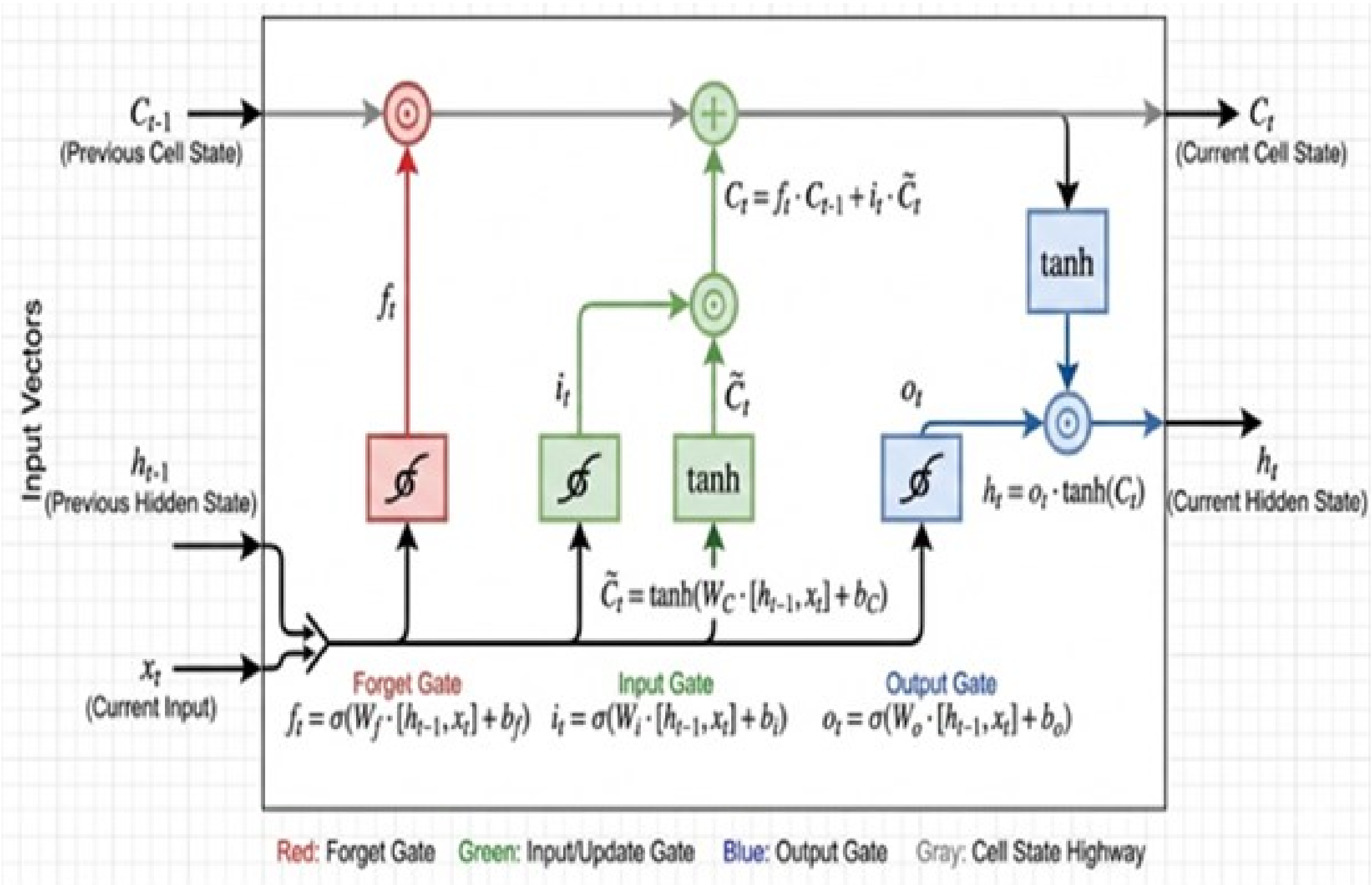

**Figure 1.** BiLSTM architecture used to classify email text.

The model was trained with binary cross-entropy loss and the Adam optimizer at a learning rate of 0.001. Training used batches of 64 for up to 20 epochs, with dropout of 0.3, recurrent dropout of 0.2, and early stopping after three epochs without improvement in validation loss. The checkpoint selected by validation F1 score was saved with its tokenizer for inference.

**Browser extension and decision rules**

A Chrome Manifest V3 extension implemented the detection workflow in Gmail. A MutationObserver detected when a message was opened, after which the extension extracted the visible sender address, subject, body text, and hyperlinks. The sender and link checks ran in the extension. Email text was sent to a Flask inference service on the same computer at 127.0.0.1:8765, which returned the BiLSTM probability.

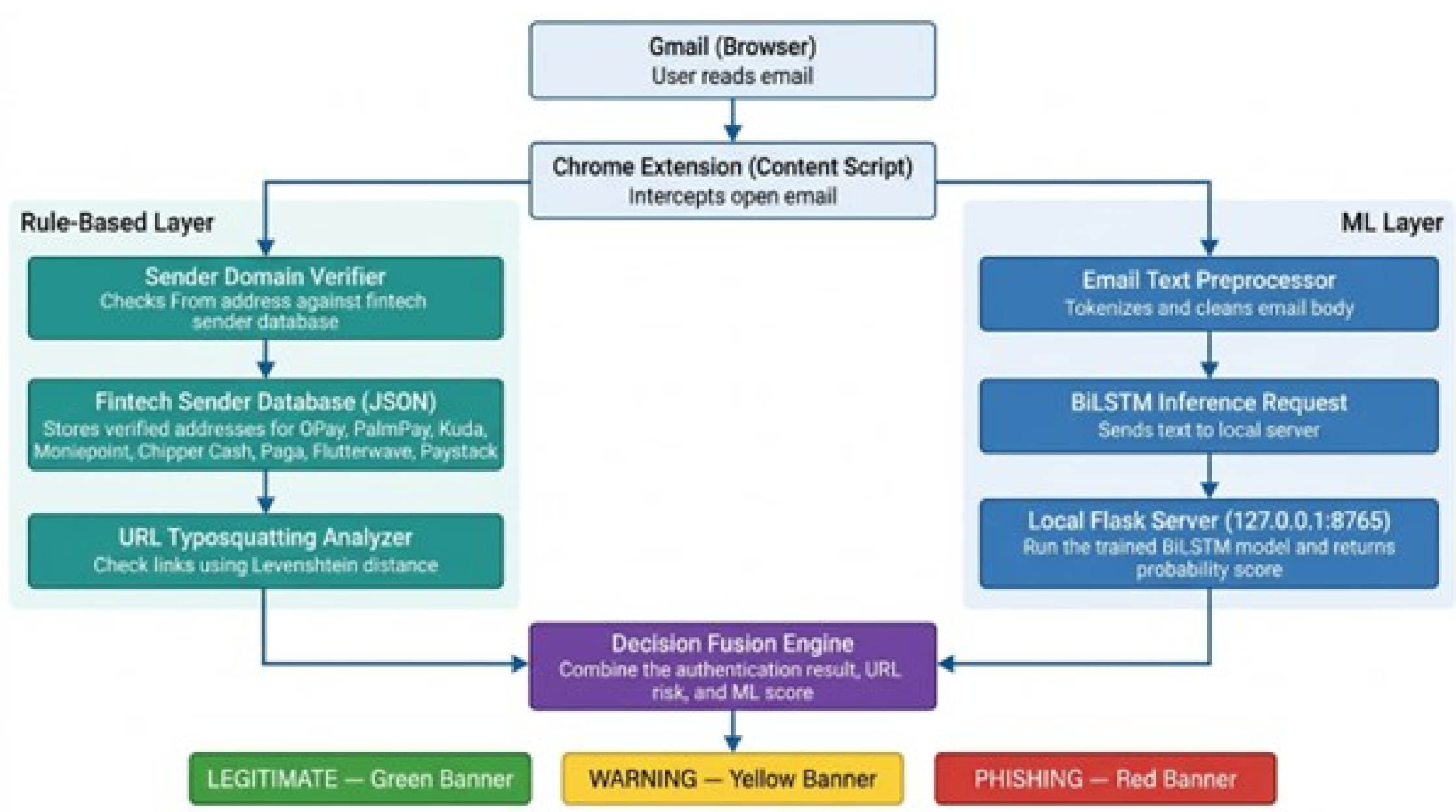

**Figure 2.** Architecture of the Gmail phishing detection extension and local inference service.

A local JSON database contained sender profiles for OPay, PalmPay, Kuda, Moniepoint, Chipper Cash, Paga, Flutterwave, and Paystack. An exact match with a listed sender domain received the internal status *VERIFIED*. A domain within a Levenshtein edit distance of two from a listed domain was marked *SUSPICIOUS*; domains without a match were treated as unverified or unknown. Extracted links were also checked for lookalike domains and Unicode homoglyphs. Here, *VERIFIED* means a match against the local profile database. The extension did not inspect raw email headers or perform SPF, DKIM, or DMARC authentication.

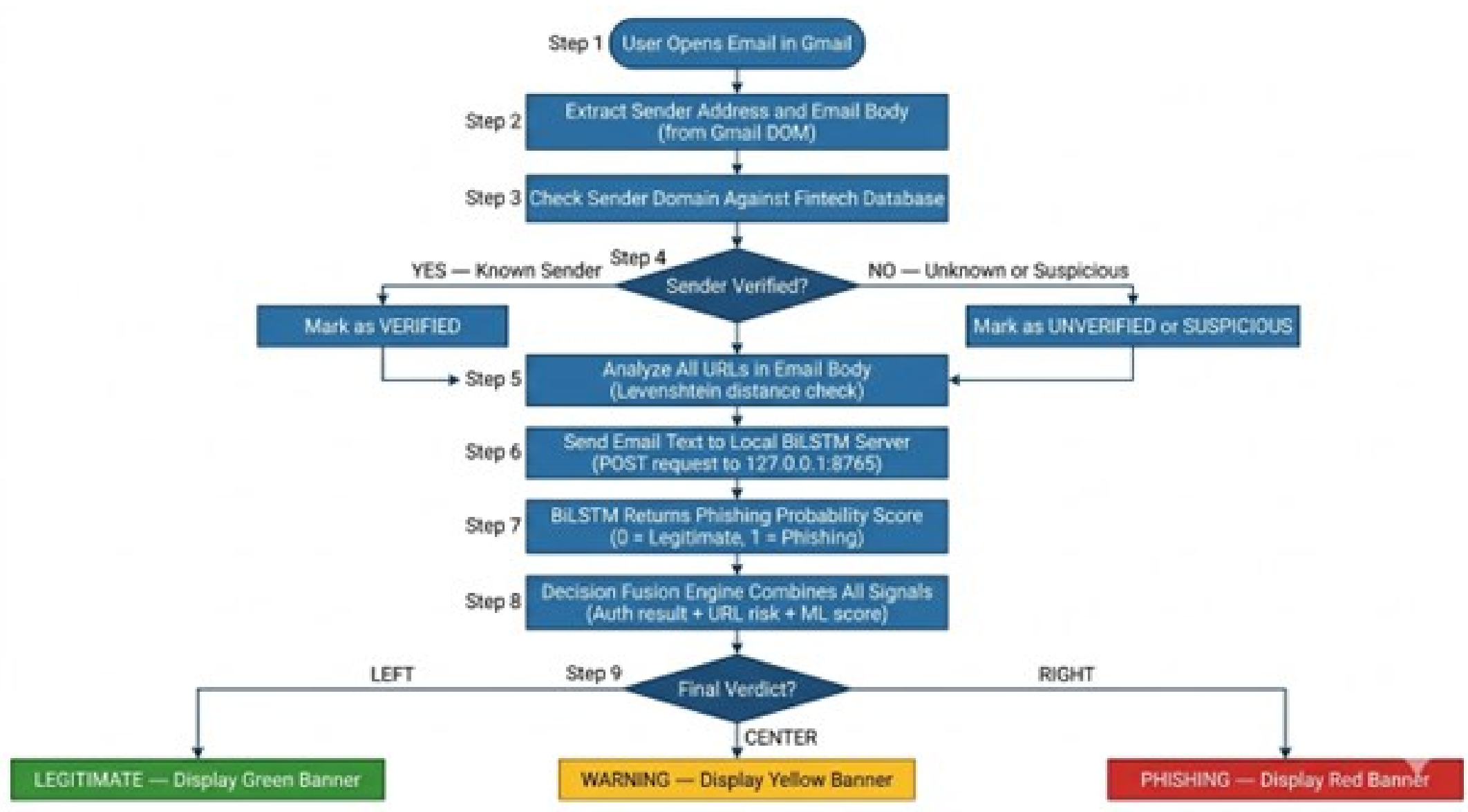

**Figure 3.** Workflow for combining sender checks, link checks, and the BiLSTM score into a verdict.

The decision rules combined sender status, link risk, and model probability into one of three displayed verdicts: LEGITIMATE, WARNING, or PHISHING. A database-matched sender with links classified as safe received a LEGITIMATE verdict regardless of the model score. A matched sender with suspicious links received a WARNING. A suspicious sender with a model score above 0.7, or an unverified sender with a score above 0.8, received a PHISHING verdict. An unverified sender with a score from 0.6 to 0.8 also received a PHISHING verdict when a suspicious link was present. Remaining cases received a WARNING. The extension displayed these verdicts as green, yellow, and red banners in Gmail.

**Evaluation**

The held-out test split was used to assess the BiLSTM classifier with a confusion matrix, accuracy, precision, recall, and F1 score. The browser extension was also exercised with selected Gmail messages to inspect extraction, rule checks, local inference, and verdict display. That demonstration did not produce a labelled sample count or a measured accuracy for the complete decision system. Accordingly, the test-set metrics reported in the next section apply to the BiLSTM classifier, not to the combined extension.

**4. Results and Discussion**

The cleaned dataset contained 59,622 emails, comprising 30,929 phishing and 28,693 legitimate messages. The BiLSTM was evaluated on the held-out test split of 8,943 emails. Its confusion matrix recorded the following outcomes:

| Actual class | Predicted legitimate | Predicted phishing |
|---|---|---|
| Legitimate | 4,308 | 0 |
| Phishing | 1 | 4,634 |

These counts yield an accuracy of 99.99%, precision of 100.00%, recall of 99.98%, and F1 score of 99.99% when rounded to two decimal places. A separate recorded metrics table gives slightly different values—99.98% accuracy, 99.97% recall, and 99.98% F1 score. The values calculated from the confusion matrix are used here because its four counts sum to the stated test-set size. The recorded ROC AUC was 1.00, but the underlying prediction scores were unavailable for independent recalculation.

An overlap check found shared tokenized sequences across the training and validation splits (5.62%), training and test splits (5.79%), and validation and test splits (1.46%). Although duplicate removal was part of data cleaning, this remaining sequence overlap raises concern about how independently the splits represent new email content. The near-perfect test result should therefore be interpreted as performance under these dataset conditions, rather than as an estimate of performance on fully independent incoming mail.

Manual checks of the rule-based layer identified lookalike sender and link domains, including examples using small spelling changes and visually similar characters. In a Gmail demonstration, the integrated extension extracted message content, obtained scores from the local inference service, and displayed all three verdict types. No labelled case count, full-system confusion matrix, or numerical latency record was retained for these checks. They establish that the components operated together, but do not establish a detection rate for the complete system.

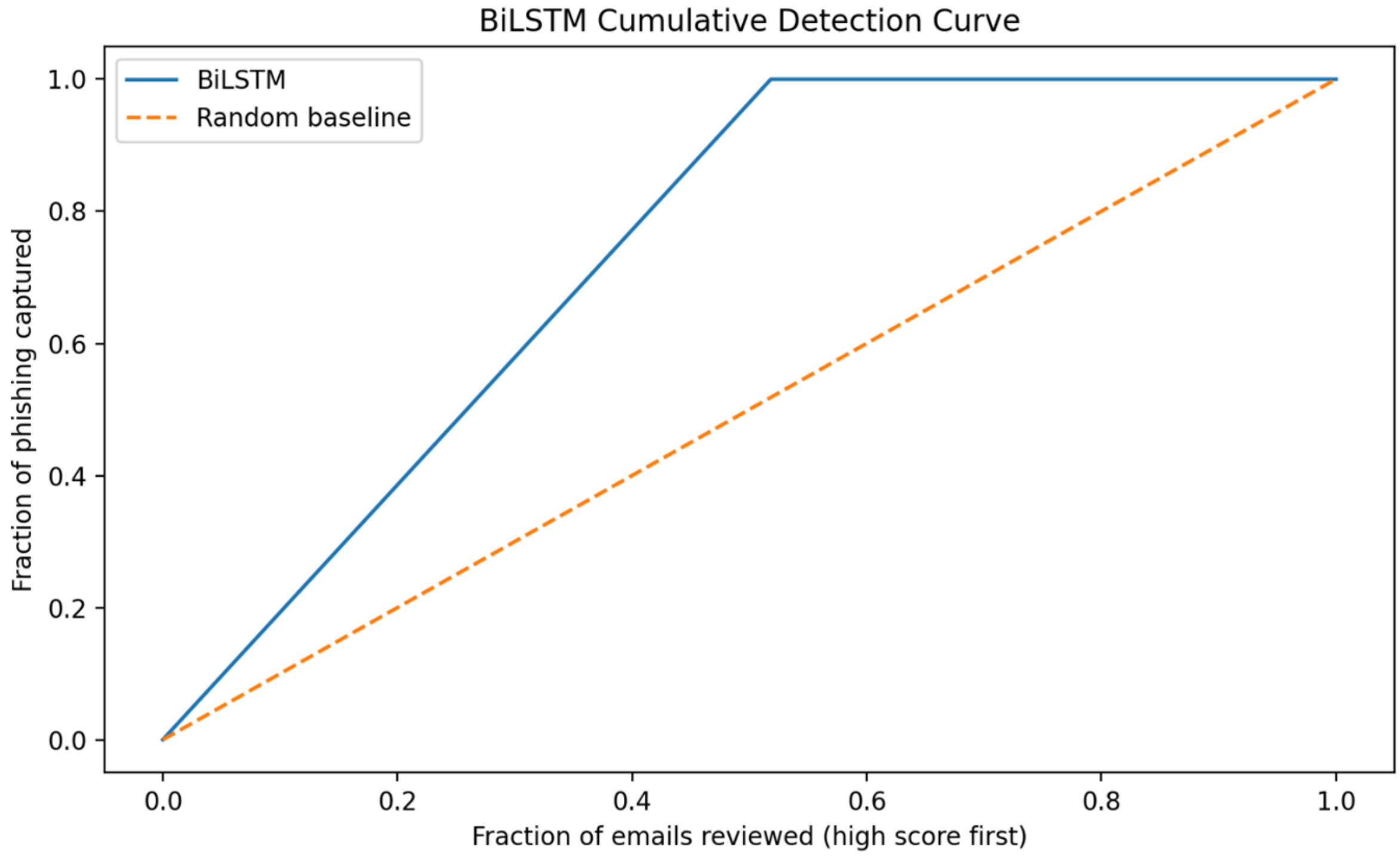

**Figure 4.** Extension verdict displayed during the Gmail demonstration.

The fusion rules also affect how the classifier's result translates into a warning. In particular, a sender whose visible domain matches a local profile receives a LEGITIMATE verdict when its links are classified as safe, regardless of the BiLSTM score. Because this profile match is not cryptographic sender authentication, the rule could suppress a model warning for a forged visible sender address. The classifier's test metrics cannot resolve that risk or be treated as metrics for the final verdict shown to users.

## 5. Limitations

The dataset limits the strength of the performance claim. Tokenized sequences overlapped across the training, validation, and test splits, so the test results may overestimate performance on unfamiliar emails. Source labels were lost during preprocessing, making it impossible to determine how many locally constructed phishing examples remained in the final dataset or to report performance separately for those examples. The available records also do not support a separate evaluation of Nigerian Pidgin messages.

The reported classification metrics describe the BiLSTM alone. The integrated extension was demonstrated in Gmail, but its verdicts were not evaluated against a counted, labelled set of messages. Consequently, the effects of the sender rules and decision fusion on false positives and missed phishing emails remain unknown.

Sender verification relied on the address visible in Gmail and a local list of known domains. It did not authenticate the sender through SPF, DKIM, or DMARC. This matters especially because an apparent match, combined with links classified as safe, overrides the model score. The local inference service also had to be running for the extension to obtain a classification, limiting use on computers where that service was unavailable.

A further evaluation should remove overlapping sequences before splitting the data, retain source and language labels, and test the complete extension on independently collected, labelled emails. That evaluation would allow separate reporting for Nigerian Pidgin messages and direct measurement of the fusion rules' effect on detection.

## 6. Conclusion

This study implemented a Gmail phishing detection extension that combines fintech sender and link checks with a BiLSTM email classifier. On a test set of 8,943 emails, the classifier produced one false negative and no false positives. The extension also demonstrated the ability to extract messages, obtain local model predictions, and display verdicts in Gmail.

The classifier's results do not establish the detection performance of the complete extension. Sequence overlap between dataset splits, the absence of a labelled evaluation of final verdicts, and reliance on visible sender addresses limit the conclusions that can be drawn. Future evaluation should use independent email samples and measure the complete decision process, including cases in which sender rules override the model.